\documentclass[aps,epsfig,floats,twocolumn,multicol,
]{revtex4}
\usepackage{amsmath}
\usepackage{amssymb}
\usepackage{graphicx}

\def\be{\begin{eqnarray}}
\def\ee{\end{eqnarray}}
\def\be{\begin{equation}}
\def\ee{\end{equation}}

\begin{document}
\title{Critical Casimir force in slab geometry with finite aspect ratio: analytic calculation above and below $T_c$}

\author{Volker Dohm}

\affiliation{Institute of Theoretical Physics, RWTH Aachen
University, D-52056 Aachen, Germany}

\date {19 February 2009}

\begin{abstract}

We present a field-theoretic study  of the critical Casimir force of the Ising universality class in a $d$-dimensional  ${L_\parallel^{d-1} \times L}$ slab geometry with a finite aspect ratio $\rho = L/L_\parallel$  above, at, and below $T_c$. The result of a  perturbation approach  at fixed dimension $d=3$ is presented that  describes the dependence on the  aspect ratio in the range $\rho \gtrsim 1/4$. Our analytic result for the Casimir force scaling function for $\rho = 1/4$ agrees well with recent Monte Carlo data  for the three-dimensional Ising model in slab geometry with periodic boundary conditions above, at, and below $T_c$.

\end{abstract}
\maketitle

In the presence of fluctuations with long-range correlations, so-called Casimir forces occur in macroscopic confined systems. The existence of such forces due to long-range critical fluctuations have been predicted  by Fisher and de Gennes \cite{fisher78} for fluid films. For such systems with isotropic interactions, the critical Casimir forces depend only on the boundary conditions (b.c.) and on the geometry of the confining surfaces as well as on the universality class of the critical point \cite{nature08,krech}. For anisotropic systems (e.g., magnetic systems with noncubic symmetry), critical Casimir forces also depend on nonuniversal anisotropy parameters \cite{cd2004,dohm2008}.

Considerable theoretical effort has been devoted to the study of critical Casimir forces in isotropic film systems over the past two decades \cite{krech92, theory, diehl06}. In the  present paper we shall focus on the Ising universality class with periodic b.c. for which detailed Monte Carlo (MC) data \cite{dan-k,vas-1} are available. While progress has been achieved by means of the $\varepsilon=4-d$ expansion  \cite{krech92,diehl06}  above the bulk critical temperature $T_c$ no theoretical prediction is available as of yet for the  region below $T_c$.  The most interesting feature is the existence of a pronounced minimum of the finite-size scaling function of the Casimir force below $T_c$ which is characteristic also for other film systems with realistic b.c. \cite{theory,vas-1,chan,hucht}.  In this Letter we present the result of a renormalization-group calculation within the framework of the $\varphi^4$ theory at fixed dimension $d=3$ \cite{dohm1985,dohm2008} that is in good agreement with the MC data \cite{dan-k,vas-1} including the minimum below $T_c$ and the Casimir amplitude at $T_c$.

All of the existing theoretical studies \cite{krech92, theory, diehl06} of the critical Casimir force in film systems have considered an $\infty^2 \times L$ geometry. This geometry is, of course, an idealization that is only approximately realized in experiments or computer simulations. In fact, the MC simulations for the Ising universality class with periodic b.c. \cite{dan-k,vas-1} have been carried out for periodic $L_\parallel^2 \times L $ slabs  with {\it finite} aspect ratios $\rho=L/L_\parallel$ in the range $1/14 \leq \rho \leq 1/3$. Most of the available data are for $ \rho = 1/6$. This appears to be well justified as the dependence on $\rho$ for  $\rho \ll 1$ is expected to be rather weak. In Ref. \cite{dan-k} it was stated explicitly that the MC results for $ \rho = 1/4$ can hardly be distinguished from those for smaller values of  $ \rho $.

Our new approach to the problem takes advantage of the fact that an $L_\parallel^2 \times L $ finite-slab geometry is conceptually simpler than a $\infty^2 \times L$ film geometry for two fundamental reasons. First, there exists no film transition at finite $\rho >0$, thus there is no necessity of dealing with the as yet unsolved problem of dimensional crossover between the 3-dimensional bulk transition and the 2-dimensional film transition. Second, for  $\rho > 0$, the system has a discrete mode spectrum with only one single lowest mode, in contrast to the more difficult situation of a lowest-mode {\it continuum} in film geometry. This opens up the opportunity of building upon the  advances that have been achieved in the description of finite-size effects in systems that are finite in all directions \cite{BZ,RGJ,dohm2008,EDC}. It is not clear {\it a priori}, however, in what range of $\rho$ such a theory is reliable since, ultimately, for sufficiently small  $\rho \ll 1$, the concept of separating a single lowest mode must break down. Therefore, as a crucial part of our theory, we first provide quantitative evidence for the expected range of applicability of our theory at finite $\rho$.

We start from the standard O$(n)$ symmetric isotropic
Landau-Ginzburg-Wilson Hamiltonian
\begin{eqnarray}
\label{1}H &=& \int\limits_{V} d^d r
\big[\frac{r_0} {2} \varphi^2 +  \frac{1} {2} (\nabla
\varphi)^2 +
 u_0 (\varphi^2)^2   \big]
\end{eqnarray}
for the $n$ component vector field $\varphi({\bf r})$ in a $d$-dimensional $L_\parallel^{d-1} \times L $ finite-slab geometry with periodic b.c. in all directions. The fundamental quantity from which the critical Casimir force  can be derived is the singular part $f_s(t,L,L_\parallel)$ of the free energy $f$ per unit volume and per component, divided by $k_B T$.  The expected asymptotic (large $L$, large $L_\parallel$, small $t=T-T_c$) finite-size scaling form of $f_s$ for isotropic systems is \cite{pri}
\begin{equation}
\label{2}f_s(t, L, L_\parallel)=L^{-d} F( x, \rho)
\end{equation}
with the scaling variable $ x=t(L/\xi_0)^{1/\nu}$ where $\xi_0$ is the amplitude of the bulk correlation length  above $T_c$.

To study the  $\rho$ dependence of the scaling function $ F( x, \rho)$ we first consider the large-$n$ limit (at fixed $u_0 n$). As an exact result we find in three dimensions
\begin{eqnarray}
\label{3} F(x, \rho) \; = \;  (8\pi)^{-1}
\left[ x P^2 \; - \; \frac{2}{3} \; P^3 \right] + \frac{1}{2}{\cal G}_0( P^2, \rho) ,
\end{eqnarray}
\begin{eqnarray}
\label{4} {\cal G}_j( P^2, \rho)= (4\pi^2)^{-j}\int\limits_0^\infty dz z^{j-1}
  \exp {\left(-\frac{P^2z}{4\pi^2}\right)}
  \nonumber\\ \times \left\{(\pi / z)^{3/2}-\Big[\rho K(\rho^2z)\Big]^2K(z)
    \right\} ,
\end{eqnarray}
with  $K(z) = \sum^{\infty}_{m= -\infty} \;\exp (- z
 m^2)$ where $P(x, \rho)$ is determined implicitly by
$P  = x  -  4\pi \;{\cal G}_1( P^2, \rho)$.

The amplitude $ F (0, \rho)$ at $T_c$ for $n=\infty$  in three dimensions is shown as thin solid line in Fig. 1. It interpolates smoothly between the  limits of $\rho=0$ (film) and $\rho=1$ (cube). It  is a monotonically decreasing function of $\rho$ since the value of $F (0, \rho)$  is suppressed  as the confinement becomes stronger. As a nontrivial feature, Fig. 1 exhibits a negligible dependence on $\rho$  for small $\rho$ up to $\rho \lesssim 1/4$. The weak dependence of $F (x, \rho)$ for $\rho \lesssim 1/4 $ also pertains to the central finite-size region $| x| \lesssim $ $O(1)$ around $T_c$. This suggests that studying $F(x, \rho)$ in a finite-slab geometry with $\rho =1/4$ should yield a good approximation to $F(x, 0)$ in film geometry near bulk $T_c$.

\begin{figure}[!h]
\includegraphics[clip,width=80mm]{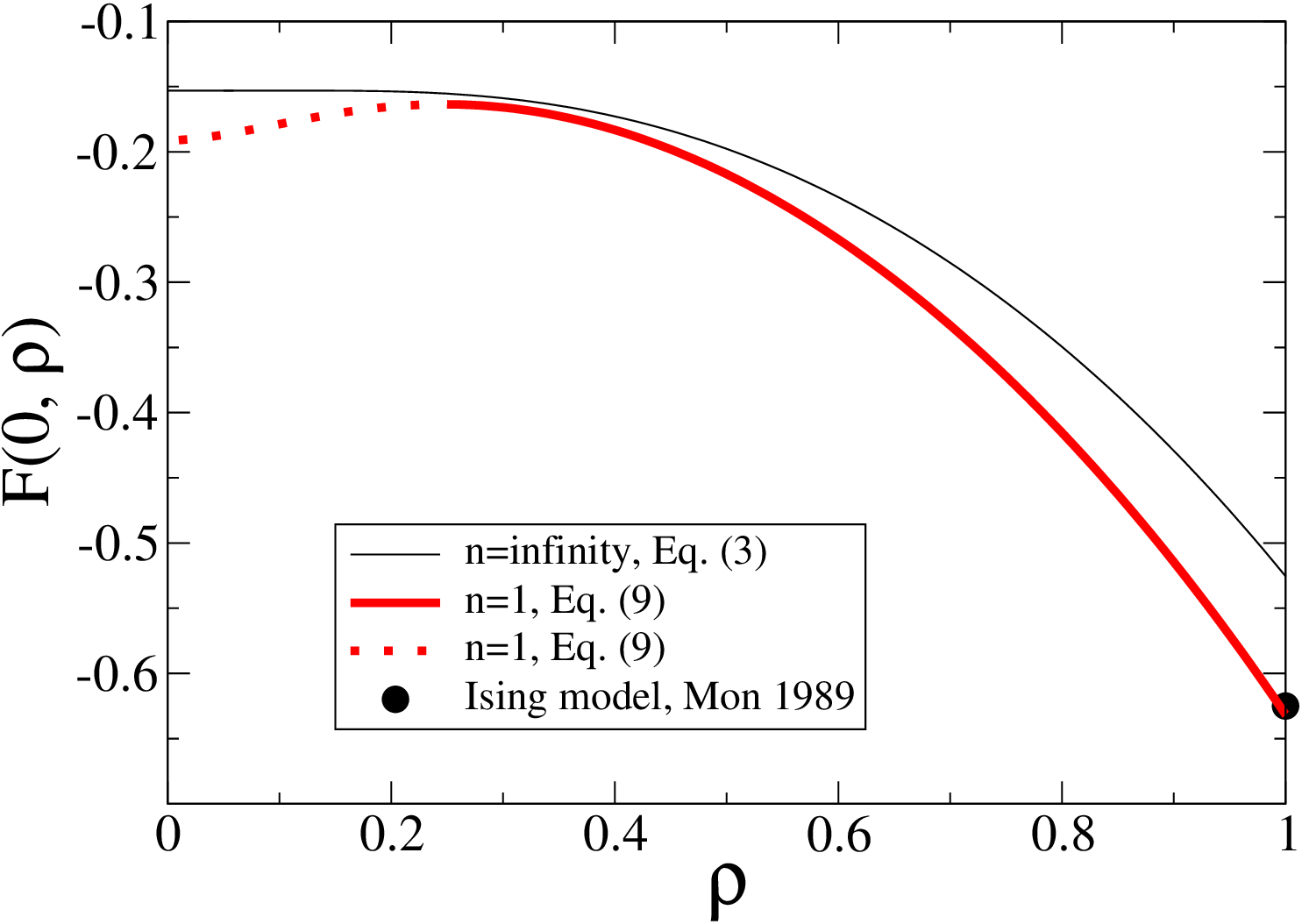}
\caption{Amplitude of the scaling function of the free energy density  $F(0, \rho)$ at $T_c$ for $n=\infty$ (Eq. (\ref{3}), thin solid line) and $n=1$ (Eq. (\ref{5}), thick solid and dotted lines) in three dimensions as a function of the aspect ratio $\rho=L/L_\parallel$. MC result (full circle) from Ref. \cite{mon} for the $d=3$ Ising model in a cube, $\rho=1$.
}
\end{figure}
One expects that a similar situation holds for $n=1$. This is indeed supported by our analytic prediction for $F(0, \rho)$ as presented in Eq. (\ref{5}) below which is  shown in Fig. 1  as thick solid and dotted lines. We have derived this result on the basis of an improved version \cite{dohm2008} of the lowest-mode separation approach \cite{BZ,RGJ,EDC}.  Our result for  $F(0, \rho)$ agrees very well with earlier MC data \cite{mon} in  a cubic geometry at $\rho=1$ (full circle in Fig. 1 ). In the range $ \rho \gtrsim 1/4$ (thick sold line), $F(0, \rho)$ has the expected negative slope.  In the range $ \rho \lesssim 1/4$, however,  the {\it positive} slope of the dotted portion of  the curve constitutes  a clear indication for the expected deterioration of the quality of the lowest-mode separation approach. In such a flat geometry with $L/L_\parallel < 1/4$ the system is already close to film geometry such that the higher modes are not well separated from the single lowest mode. On the other hand, together with the result for $n=\infty$, Fig. 1 suggests that a calculation of $F(x, \rho)$  for $n=1$ at $\rho= 1/4$ should yield an acceptable approximation to $F(x, 0)$ in film geometry near bulk $T_c$.

Our derivation of $F(x, \rho)$ for $n=1$ is based on the Hamiltonian (\ref{1}) where the decomposition $\varphi = \Phi + \sigma $ is made into a  homogeneous lowest-mode amplitude $\Phi$ and  higher-mode fluctuations $\sigma$. After integration over $\sigma$, the free energy density is obtained in the form
\be
\label{6}
f = f_0  -V^{-1}\ln \int\limits_{-
\infty}^\infty d \Phi \exp \left[-H_0(\Phi) - \Gamma(\Phi)\right]
\ee
with $H_0(\Phi)=V(\frac{1}{2}r_0\Phi^2 + u_0\Phi^4)$ where $f_0$ is independent of $r_0$ and $u_0$. The higher-mode contribution $\Gamma(\Phi)$ is calculated in one-loop order  and  expanded around the lowest-mode average $M_0^2= \int d\Phi \Phi^2 e^{-H_0}/\int d\Phi  e^{-H_0}$ up to  $ O((\Phi^2- M_0^2)^2)$.  In truncating our expansion of $\Gamma(\Phi)$ we require that, in the central finite-size region including $T=T_c$,  terms of $ O (u^{3/2}_0)$ are neglected. As we are working at fixed dimension $2<d<4$ there is no necessity of further expanding the exponential function $e^{-H_0-\Gamma}$. Thus we maintain the exponential structure of the integrand in (\ref{6}). The resulting bare perturbation expression for $f$ contains the bare bulk free energy density $f_b^\pm \equiv \lim_{V\rightarrow \infty}f$ in one-loop order above (+) and below ($-$) $T_c$. The dependence of $f - f_b^\pm$ on the aspect ratio $\rho$ appears (i) on the level of the lowest-mode Hamiltonian $H_0(\Phi)$ and (ii) on the level of the contribution of $\Gamma(\Phi)$. The former dependence (i) comes from $M_0^2(r_0,L,\rho)= (L^d\rho^{1-d}u_0)^{-1/2}\vartheta(y_0)$ with $y_0= r_0(L^d\rho^{1-d}/u_0)^{1/2}$ where
\be
\vartheta (y) =\int\limits_0^\infty d z z^2e^{-
\frac{1}{2} y z^2 - z^4}/\int\limits_0^\infty d z e^{-
\frac{1}{2} y z^2 - z^4}.
\ee
The latter dependence (ii) is contained in the difference between sums over higher modes and bulk integrals in wave vector $(\mathbf k)$ space such as
\begin{eqnarray}
\label{c3}  V^{-1} \sum_{{\bf k} \neq {\bf 0}} (r_0 +
 \mathbf k^2) ^{-m} -
\int\limits \frac{d^d\mathbf k}{(2\pi)^d}(r_0 + \mathbf k^2)^{-m} = \nonumber\\
\frac{1-\rho^{d-1}}{L^d} r_0^{-m}+ \frac{L^{2m-d}}{(4 \pi^2)^m}
I_m (r_0L^2, \rho) ,
\end{eqnarray}
\begin{eqnarray}
\label{c4} I_m (r_0L^2, \rho) = \int\limits_0^\infty
{\rm{d}}z \;z^{m-1} \exp[- r_0L^2 z / (4 \pi^2)] \nonumber\\
\times \Big\{\Big[\rho K(\rho^2z)\Big]^{d-1}K(z) - (\pi / z)^{d/2} - 1\Big\}
\qquad.
\end{eqnarray}
The bare perturbation result needs, of course, to be renormalized. Within the minimal renormalization scheme in three dimensions  \cite{dohm1985} we have obtained the  following scaling function for $n=1$
\begin{eqnarray}
\label{5} &&F( x, \rho)= - \;
\frac{ l^3}{48 \pi} \; - \;\frac{\nu\;{Q^*}^2  x^2
l^{- \alpha/\nu}}{16 \pi \alpha} + \frac{1}{2}{\cal G}_0(l^2, \rho)\nonumber\\ && + \; 18
u^*\rho^2 \left[\vartheta ( y)\right]^2  + ( \rho^2 - 1)[ a(x,\rho) +a(x,\rho)^2]\nonumber\\
&& - b(x,\rho)\; I_1( l^2, \rho) - b(x,\rho)^2
\; I_2( l^2, \rho)\nonumber\\&&-\;\rho^2  \ln \int\limits_{-\infty}^\infty d z \;\exp \big[-
\frac{1}{2}  Y( x, \rho) z^2 - z^4\big]  \nonumber\\&&
\; -   \frac{\rho^2}{2}\ln \left\{\frac{ { l}^{3/2}\left[1 + 18 \; u^* R_2(  l, \rho) \right]}{4 {u^*}^{1/2}\pi^{3/2} \rho}\right\}
\end{eqnarray}
where
\begin{eqnarray}
\label{6i} &&  Y( x, \rho
) = { l}^{ 3/2}\rho^{-1}  (4\pi{u^*})^{-1/2} \Bigg\{24 u^* a(x,\rho)R_2( l, \rho) \nonumber \\&& + Q^*  x \;{ l}^{-1/\nu} \Big[1 + 18
u^*R_2( l,\rho) \Big]
 +  12 u^* R_1(l,\rho)  \Bigg\}
\end{eqnarray}
with $a(x,\rho)=12
{u^*}^{1/2}  l^{-3/2}\rho\pi^{1/2}
\vartheta( y)$ and $b(x,\rho)=3{
l}^{1/2} {u^*}^{1/2}\rho \pi^{-3/2} \vartheta(y)$, $u^*=0.0412, Q^*=0.945, \nu=(2-\alpha)/3=0.6335$.
The functions   $l(x,\rho)$, $y(x,\rho)$, and $R_i(l,\rho)$ are determined by
\be
\label{6f}  y + 12 \vartheta(y) = \rho^{-1}
 l^{3/2} (4\pi {u^*})^{-1/2},
\ee
\be
\label{6g}  y =\;  x\;Q^*\; l^{- \alpha / (2
\nu)}\rho^{-1} (4\pi {u^*})^{-1/2},
\ee
\begin{eqnarray}
R_1(l,\rho) =
4\pi (1 - \rho^2)  l^{-3}
+  (l\pi )^{-1} I_1 (l^2,\rho),
\end{eqnarray}
\begin{eqnarray}
R_2 ( l, \rho) =
- \frac{1}{2}  + 4\pi (1 - \rho^2)
 l^{-3}
+ \; (l /4\pi^3) I_2 (l^2, \rho) \; .
\end{eqnarray}
For finite  $\rho > 0$, $F( x, \rho)$ is
an analytic function of $x$ near $x = 0$, in agreement with
general analyticity requirements.
\begin{figure}[!h]
\includegraphics[clip,width=80mm]{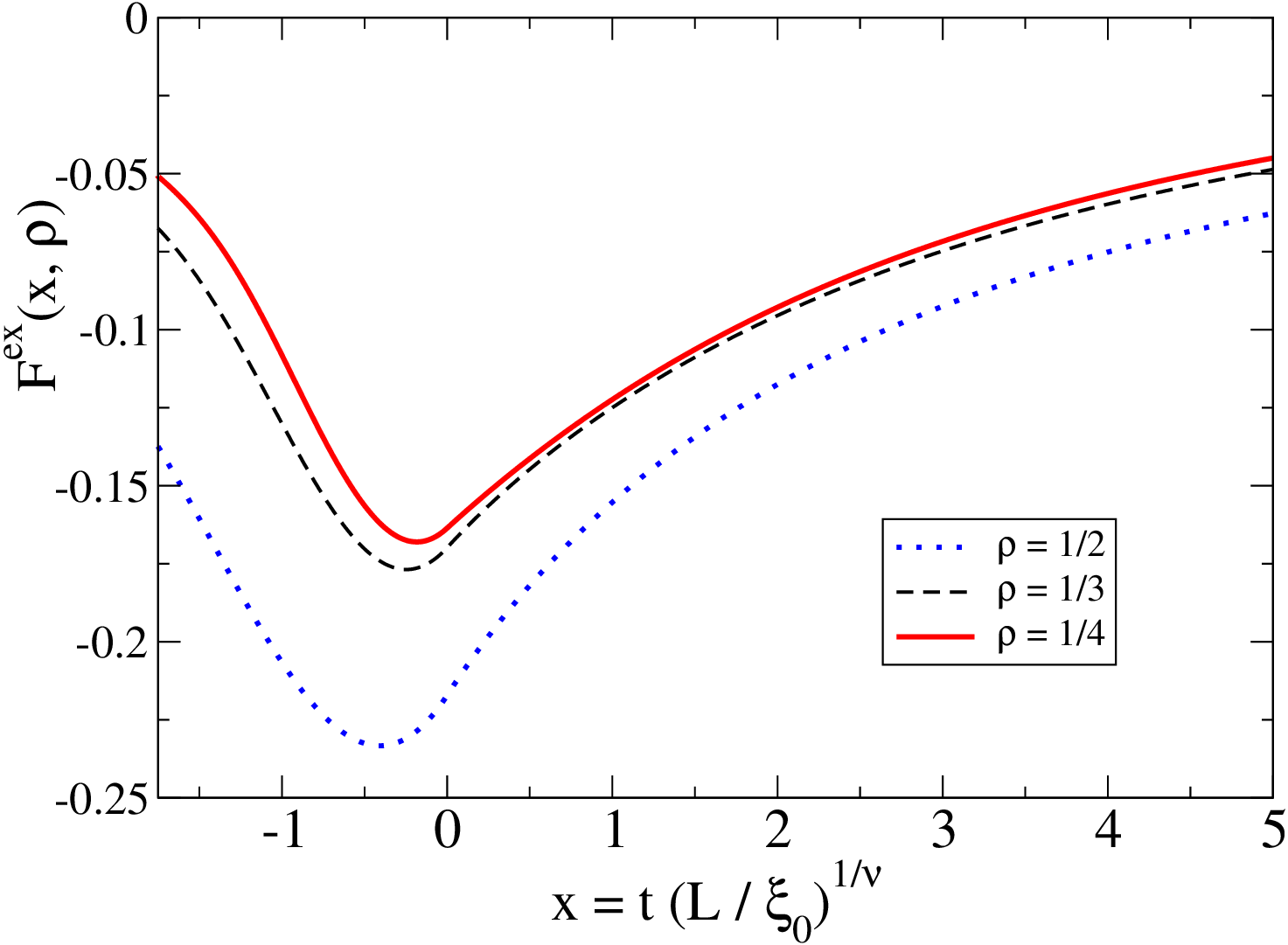}
\caption{Scaling function $F^{ex}(x, \rho)$  of the excess free energy density  for $n=1$, Eqs. (\ref{5}), (\ref{3p}), (\ref{14}) for $\rho=1/2, 1/3, 1/4$  in three dimensions as a function of the scaling variable $x$. }
\end{figure}

The singular part of the bulk free energy density $f^\pm_{s,b}(t)=A^\pm |t|^{d\nu}$ is, of course, independent of $\rho$. It can be written as $f^\pm_{s,b}= L^{-d}{F}^\pm_b(x)$ where ${F}^\pm_b(x)$ is the bulk part of $F(x, \rho)$  which is obtained from (\ref{5}) in the limit of large $|x|$. It is given by
\begin{eqnarray}
\label{3p}
{F}^\pm_b(x) =\left\{
\begin{array}{r@{\quad \quad}l}
                         \; Q_1  x^{d \nu}\quad          & \mbox{for} \;T > T_c\;, \\
                         \; (A^-/A^+)Q_1\mid x\mid^{d\nu}& \mbox{for} \;T <
                 T_c \;,
                \end{array} \right.
\end{eqnarray}
with universal numbers $Q_1=-0.119$ and $A^-/A^+ = 2.04$ in three dimensions. Thus the singular part of the excess free energy density $f_s^{ex}= f_s - f^\pm_{s,b}$ has the scaling form
\be
\label{13}
f_s^{ex}(t,L,L_\parallel)= L^{-d}F^{ex}(x,\rho),
\ee
\be
\label{14}
F^{ex}(x,\rho)= F(x,\rho) - {F}^\pm_b(x).
\ee
The prediction of the  $\rho$ dependence  of $F^{ex}(x, \rho)$ above, at, and below $T_c$  for the Ising universality class as described by (\ref{5})-(\ref{14}) without any adjustment of parameters is the central result of this paper. This function contains a $\rho$ dependent minimum slightly below $T_c$. The scaling function $F^{ex}(x,\rho)$ is shown in Fig. 2 for several values of $\rho$. As expected on the basis of Fig. 1, the  difference between  $F^{ex}(x,\rho)$ for $\rho=1/3 $ and $\rho=1/4 $ is rather small.

\begin{figure}[!h]
\includegraphics[clip,width=80mm]{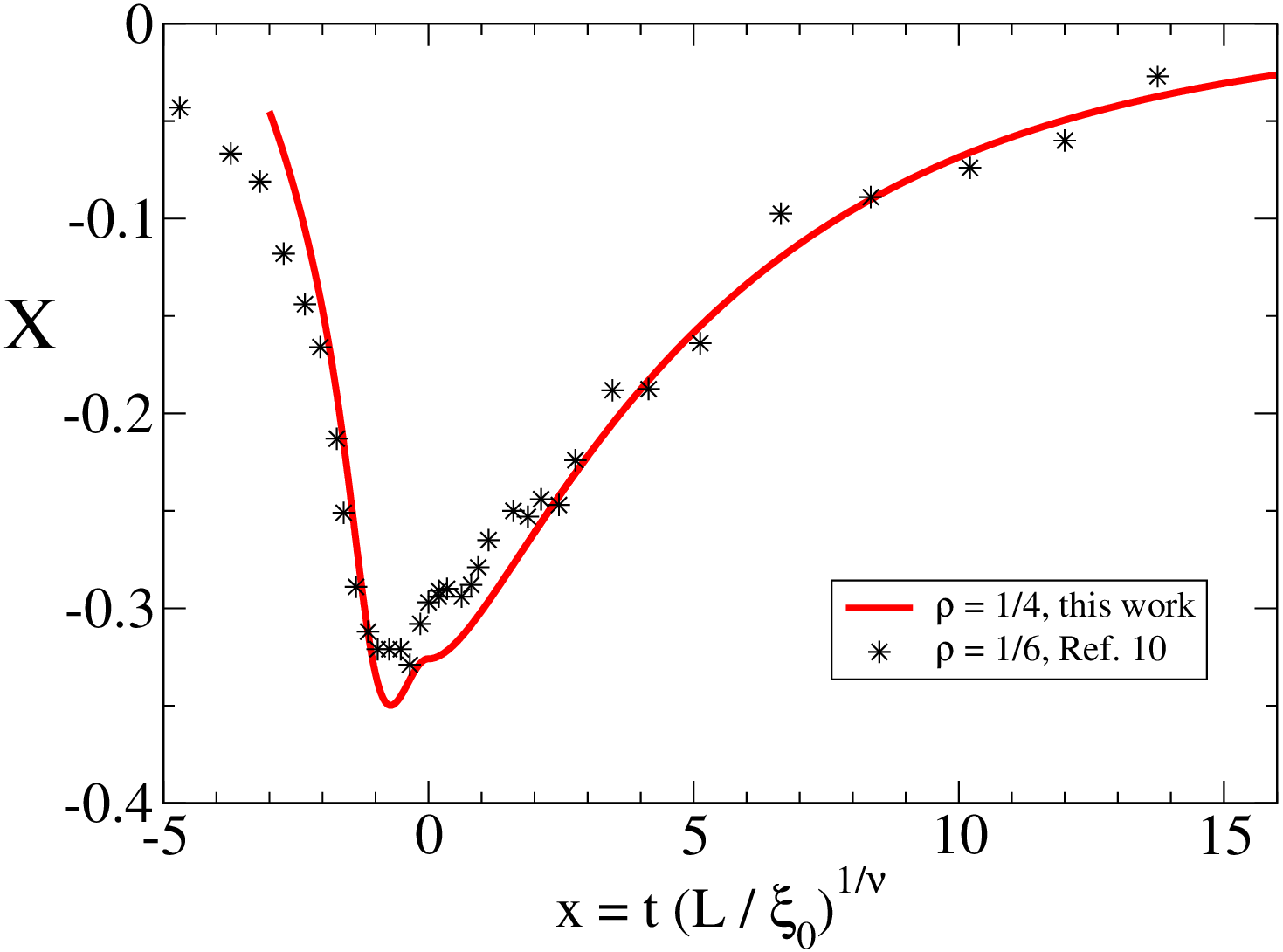}
\caption{Scaling function $X(x,\rho)$, Eqs. (\ref{5}), (\ref{3p}), (\ref{14}), (\ref{3n})  of the Casimir force  in three dimensions
for finite-slab geometry:  $ \varphi^4$  theory at $d=3$  (this work, solid line) for $\rho=1/4$, MC data  \cite{vas-1} (stars) for $\rho=1/6$.}
\end{figure}

We note that so far no confirmation of the theory at $\rho=1$ \cite{dohm2008} by MC simulations has been presented  except right at $T=T_c$ \cite{mon}. In particular the prediction  of a minimum of $F(x, 1)$  below $T_c$ for $\rho=1$ \cite{dohm2008} is as yet unconfirmed since no MC data are available as of yet in this regime. For this reason it is particularly interesting to present here our prediction of a minimum of the Casimir force scaling function for small $\rho$ slightly below $T_c$ which can be compared  with recent MC data \cite{dan-k,vas-1}.

We define the critical Casimir force $F_{Casimir}$ per unit area in a finite-slab geometry as
\be
F_{Casimir}(t,L,L_\parallel)=-\frac{\partial [L f_s^{ex}]}{\partial L}=L^{-d}X(x,\rho)
\ee
where the derivative is taken {\it at fixed} $L_\parallel$. This definition is equivalent to its lattice counterpart introduced in \cite{vas-1}. The Casimir force scaling function $X$ can then be expressed in terms of $F^{ex}$ as
\begin{equation}
\label{3n} X(x,\rho) =(d-1)  F^{ex} (x,\rho) -
\frac{x}{\nu}
\frac{\partial{ F^{ex} }(x,\rho)}{\partial x}- \rho
\frac{\partial{ F^{ex} }(x,\rho)}{\partial \rho}.
\end{equation}
\begin{figure}[!h]
\includegraphics[clip,width=80mm]{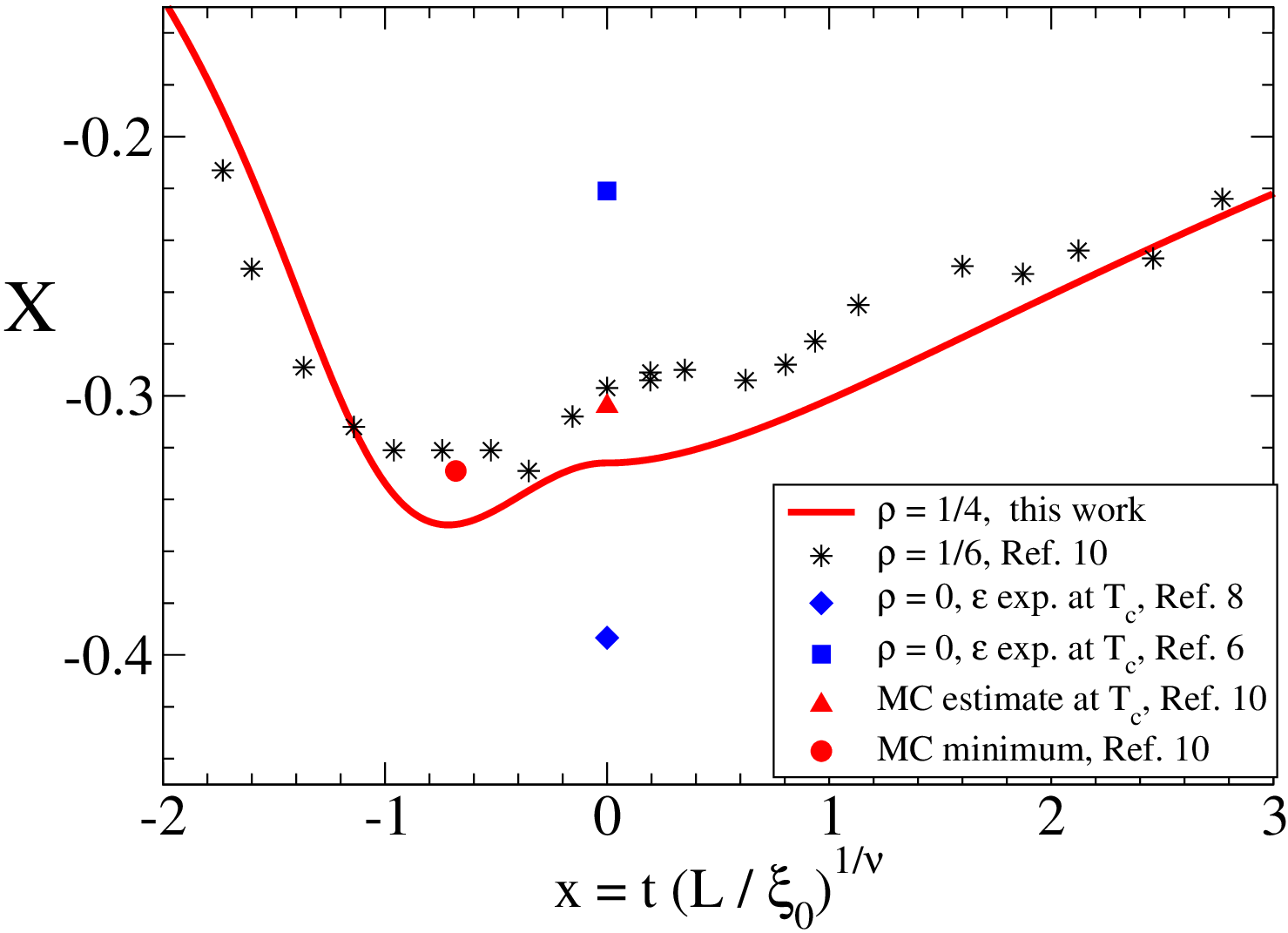}
\caption{Magnified plot of Fig. 3 near $T_c$: MC estimate for $X^{MC}_c$ at $T_c$ (triangle) \cite{vas-1}, MC estimate for the minimum $X^{MC}_{min}$ (circle) \cite{vas-1}, two- and three-loop $\varepsilon$ expansion results at $T_c$ (square and diamond) for $\rho=0$ \cite{krech92,diehl06}.}
\end{figure}

Numerical evaluation of (\ref{3n}) yields the curve shown in Fig. 3. There is good overall agreement with the MC data  of  \cite{vas-1} (with $\rho = 1/6$ and $L=20$) and with the MC data of \cite{dan-k} (not shown in Fig. 3) in the range $-2 \lesssim x \lesssim 15$. Somewhat unexpectedly, our result exhibits a small shoulder near $T_c$. This shoulder is not  present in the scaling function of the excess free energy density $F^{ex}(x, \rho)$, Fig. 2, but arises through the derivative term $ -(x/\nu) \partial F^{ex} (x,\rho)/\partial x$.

A more detailed comparison with earlier results is shown in Fig. 4. Most significant is the satisfactory agreement of the position  of the minimum of the theoretical curve
$x_{min}= -0.715$ with the MC estimate \cite{vas-1} $x^{MC}_{min}= -0.681$ (full circle in Fig.4). There is also reasonable agreement with regard to the  depth of the theoretical minimum  $X(x_{min},1/4)= -0.350$  compared to the MC estimate \cite{vas-1}   $X^{MC}_{min}= -0.329$ (full circle in Fig.4). Furthermore, our result $X(0,1/4) = -0.326$ at $T_c$ is in substantially improved agreement with the MC estimate \cite{vas-1} $X^{MC}_c=-0.304$  at $T_c$ (triangle in Fig. 4), compared to the earlier $\varepsilon$ expansion results $-0.221$ in two-loop order \cite{krech92} and  $-0.393$ in three-loop order \cite{diehl06} (shown in Fig. 4 as square and diamond, respectively).

In summary, we have presented a new approach to the analytic calculation of the critical Casimir force scaling function in slab geometry for isotropic systems in the Ising universality class and have obtained quantitative agreement with MC data  for periodic boundary conditions. This approach can be extended to realistic boundary conditions and to other universality classes which may then lead to a satisfactory explanation of the minimum of the critical Casimir force scaling function below $T_c$ in real systems \cite{chan}.

\end{document}